# Country wide Shared FibreBased Infrastructure for Dissemination of Precise Time, Coherent Optical Frequency with Vibration Sensing

Josef Vojtech[a], Tomas Novak[a,d], Elisabeth Andriantsarazo[a,c], Vladimir Smotlacha[a], Ondrej Havlis[a], Rudolf Vohnout[a], Michal Spacek[a,c], Martin Slapak[a], Lada Altmannova[a], Radek Velc[a], Petr Pospisil[a], Jan Kundrat[a], Martin Cizek[b], Jan Hrabina[b], Simon Rerucha[b], Lenka Pravdova[b], Josef Lazar[b], Ondrej Cip[b], Jaroslav Roztocil[c]

[a]CESNET z.s.p.o., Prague, Czechia; [b]Institute of Scientific Instruments of the Czech Academy of Sciences, Brno, Czechia; [c]Faculty of Electrical Engineering, Czech Technical University, Prague Czechia; [d]Faculty of Nuclear Science and Physical Engineering, Czech Technical University, Prague Czechia

**ABSTRACT**

With the increasing demand for ultra-precise time synchronization and frequency dissemination across various scientific, industrial, and communication fields, the Czech Republic has developed an innovative, non-commercial fiber-based infrastructure. This infrastructure serves as a shared platform, utilizing optical fibers to enable high-precision timing, coherent frequency transfer, and a newly implemented vibrational sensing capability. The project also addresses challenges posed by classical communication noise—particularly from Raman scattering—on quantum channels, especially for Quantum Key Distribution (QKD). By strategically separating classical and quantum channels into distinct wavelength bands, such as the C-band and O-band, the infrastructure achieves minimal interference while enabling multiple concurrent applications over shared fiber lines.

## 1. INTRODUCTION

In modern networking and metrology, synchronization and stable frequency transmission are paramount. Applications such as metrology, navigation, Earth sciences, seismology, and quantum communication rely on highly precise timekeeping and frequency control to support their operations. Traditional systems, such as those utilizing GPS, cannot deliver the ultra-stable parameters required for these advanced fields. As a result, fiber-optic infrastructures have emerged as a superior alternative, offering high precision and stability over long distances.

The Czech Infrastructure for Time and Frequency (CITAF) has pioneered non-commercial fiber-based systems dedicated to time and frequency dissemination across national and international networks [1]. This project, driven by CESNET (Czech Education and Scientific NETwork) association and the Czech Academy of Sciences, utilizes the CESNET3 network, a shared optical fiber infrastructure. Building on CITAF's foundations, the current project introduces vibrational sensing as a new dimension for monitoring environmental changes along fiber routes and ensures secure quantum-safe communication through QKD. This paper discusses the technical approach, challenges in mitigating Raman scattering, and the strategies implemented to maintain the fidelity of classical and quantum channels over shared fibers.

## 2. SHARED FIBER INFRASTRUCTURE FOR TIME AND FREQUENCY TRANSMISSION

Optical fibers are increasingly used to transmit ultra-stable time and frequency signals over long distances. CESNET has developed an infrastructure that incorporates these transmissions in shared fiber networks, implementing both time synchronization protocols and frequency transfer methods. Since its inception in 2010, CESNET's infrastructure has expanded to connect neighboring countries, providing transnational precise time and frequency dissemination [2,3].

To achieve the highest levels of precision, CITAF deploys White Rabbit (WR) technology, an advanced protocol that synchronizes clocks with sub-nanosecond precision over Ethernet. WR employs Synchronous Ethernet (SyncE) and IEEE 1588 Precision Time Protocol (PTP), utilizing two-way message exchanges to ensure precise clock phase and offset synchronization [4]. Dedicated channels in the C-band, operating around 1550 nm, are allocated for these transmissions, offering stable, low-noise conditions essential for high-precision applications.

The CESNET association for the transmission of ultra-stable quantities of precise time and coherent frequency most commonly started to use a portion of the reserved optical spectrum in the C-band in the range of 1540–1546 nm (8 channels). Unfortunately this allocation represents a problem for modern spectrally flexible broadband data transfers, thus also a portion of the optical spectrum at the between the C and L bands in the range of 1570–1572 nm (4 channels) has been reserved and started to be used [5]. In recent research activities, the possibility of using the S-band at a wavelength of 1458 nm was also verified [6, 7]. Overall spectrum arrangement is shown in Fig. 1.

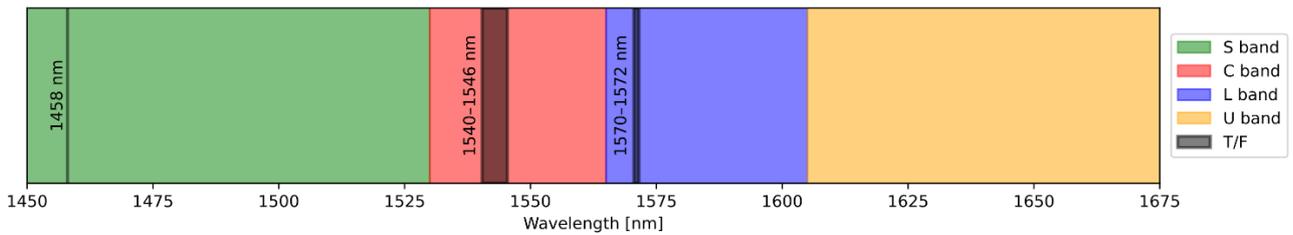

Figure 1. Spectrum allocation for ultrastable quantities transfer

## 3. RAMAN SCATTERING AND WAVELENGTH SEPARATION FOR CLASSICAL AND QUANTUM CHANNELS

The intended simultaneous transmission of classical and quantum signals in shared fiber infrastructure introduces unique challenges. In particular, the nonlinear phenomenon known as spontaneous Raman scattering, resulting from inelastic scattering of photons, can introduce noise in quantum channels. When classical communication channels transmit time and frequency data, Raman scattering generates noise photons that interfere with the QKD measurement process, potentially jeopardizing the integrity of quantum communications.

In Raman scattering, a photon's energy is altered through interactions with molecular vibrations within the optical fiber, leading to either Stokes or anti-Stokes shifts in frequency. The scattered photons retain the polarization of the original signal, complicating the isolation of noise from the signal in quantum communication channels. Measurements using band-pass filters reveal that the power of Raman noise from classical channels operating in the C-band can reach levels comparable to the quantum signals, which may reach ~$10^5$ photons per second in intensity, see Fig.2.

We can calculate the approximate amount of anti-Stokes Raman-scattered photons generated from the input signal at 1550 nm to 1320 nm using the theory of Raman scattering. This involves considering the spontaneous anti-Stokes Raman scattering process, which is inherently weaker than the Stokes process due to the lower population of higher-energy vibrational states at room temperature. The Raman-scattered light has a shorter wavelength (higher energy) when it gains energy from thermally excited vibrations in the fiber. The efficiency of spontaneous Raman scattering is relatively low but can be significant over long fiber lengths or at high input powers. Attenuation reduces both the input signal and the scattered light over distance, leading to a significant impact on overall performance.

The anti-Stokes scattered power is extremely small compared to the input power of the classical signal. The large frequency shift (33.72 THz) places the anti-Stokes wavelength far from the Raman gain peak, further reducing the gain coefficient. When assessing potential interference or crosstalk in fiber-optic systems, anti-Stokes scattering is often considered negligible unless operating conditions (e.g., high temperatures or specialized materials) significantly alter the thermal population of vibrational states. However, this level of noise is still significant for quantum communication systems and must be addressed.

To mitigate the interference caused by Raman scattering, a wavelength separation strategy must be implemented. Operating strong channels (also amplified coherent optical frequency and time transfers, where massive Amplified Spontaneous Emission is present [8]) brings the necessity to use very low wavelengths for quantum channels (<1290 nm), which suffer from excessive attenuation. Optical filters, including Coarse Wavelength Division Multiplexing (CWDM) and Dense Wavelength Division Multiplexing (DWDM), are employed to maintain this separation. These filters are critical for ensuring that the various signals do not interfere with each other, thus suppression of unwanted signal over the whole telecom band, low insertion loss, and high return loss are preferred, more details in [9].

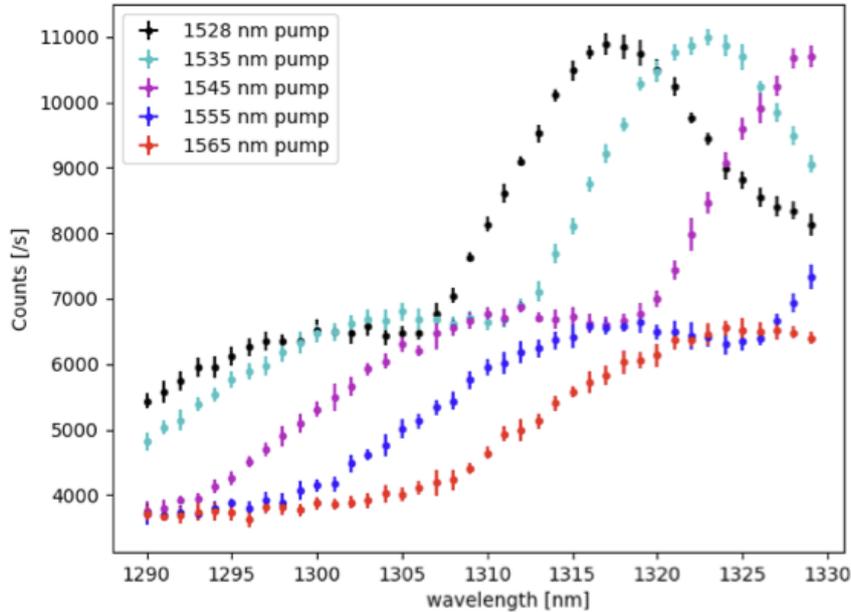

Fig. 2: The Raman gain spectrum from five different pump wavelengths.

### 4. PROSPECTIVE APPLICATION OF VIBRATIONAL SENSING

This infrastructure also includes a vibrational sensing option, which allows monitoring of environmental changes in real-time. This capability enables detection of physical disturbances or structural changes along the fiber route, enhancing the system's utility for applications beyond metrology. There are multiple possibilities, the most affordable ones are represented by utilizing Doppler noise canceling processes on lines with coherent frequency transfers [10] or polarimetry [11]. Such a capability would be valuable for applications like detecting seismic activities, and ensuring security by identifying physical disturbances or potential breaches along the fiber route. This addition would further enhance the CITAF infrastructure's versatility, offering an extra layer of utility for scientific, security, and environmental monitoring purposes.

### 5. CONCLUSIONS AND FUTURE PROSPECTS

The Czech Republic's nationwide, shared fiber infrastructure for precise time and coherent optical frequency dissemination demonstrates an effective, cost-efficient approach to supporting diverse applications in a single network. By carefully addressing the challenges posed by Raman scattering through spectral separation and advanced filtering, this infrastructure maintains the fidelity of quantum communications while facilitating high-precision time and frequency transfers. Future developments will expand this network's reach across Europe through the implementation of some project CLONETS-DS proposals [12], aimed at creating a continental clock network. It is realized within the C-TFN activity of GÉANT project GN5-1 [13]. With its multi-functional capabilities—including precise time transfer, coherent frequency dissemination, vibrational sensing, and quantum-safe communication—this infrastructure is poised to support a new era of interconnected, ultra-precise technologies for scientific and industrial applications.

## ACKNOWLEDGEMENTS

This work was supported partially by The Ministry of Education, Youth, and Sports of the Czech Republic by project EH22_008/0004649 Quantum Engineering and Nanotechnology and partially by The Ministry of Interior of the Czech Republic through NU-CRYPT - Quantum encrypted communication with increased physical layer security (VK01030193).

## REFERENCES

[1] https://citaf.org/ (Accessed: 1. August 2024).
[2] V. Smotlacha, A. Kuna, W. Mache: Time Transfer in Optical Network. Proc. 42nd Annual Precise Time and Time Interval (PTTI) Systems and Applications Meeting, Reston, Virginia, USA, (U.S. Naval Observatory, Washington D.C.), (2010) 427–36.
[3] V. Smotlacha, A. Kuna, W. Mache: Optical Link Time Transfer between IPE and BEV. Proc. 43rd Precise Time and Time Interval (PTTI) Systems and Applications Meeting (2011) 27–34
[4] Lipinski, M., Wlostowski, T., Serrano, J., and Alvarez, P., "White rabbit: A ptp application for robust sub-nanosecond synchronization," IEEE International Symposium on Precision Clock Synchronization for, Measurement, Control, and Communication, ISPCS , 25–30 (2011).
[5] Josef Vojtech, et al , "Joint accurate time and stable frequency distribution infrastructure sharing fiber footprint with research network," Opt. Eng. 56(2) 027101 (6 February 2017) https://doi.org/10.1117/1.OE.56.2.027101
[6] Vojtech, Josef, et al ,"Alternative Spectral Window for Precise Time Fibre Based Transport," Proceedings of the 51st Annual Precise Time and Time Interval Systems and Applications Meeting, San Diego, California, January 2020, pp. 187-190. https://doi.org/10.33012/2020.17313
[7] Josef Vojtech, et al , "Optical amplification for quantum sources of ultra-stable optical frequency", Proc. SPIE 10976, 21st Czech-Polish-Slovak Optical Conference on Wave and Quantum Aspects of Contemporary Optics, 109760Q (18 December 2018); https://doi.org/10.1117/12.2513641.
[8] M. Slapak, et al, "Stabilization of super coherent frequency transfers via amplifier cascade balancing", Optical Fiber Technology 87, (2024). https://doi.org/10.1016/j.yofte.2024.103910
[9] J. Radil, et l, "Quantum and Data Signals in a Single Fibre –Multiplexing by Smart Use of High-Grade Filters" submitted into IEEE Communications letters.\
[10] Čížek, M. et al "Coherent fibre link for synchronization of delocalized atomic clocks. Optics Express. 30. 10.1364/OE.447498.
[11] M. Šlapák, J. Vojtěch, O. Havliš and R. Slavík, "Monitoring of Fibre Optic Links With a Machine Learning-Assisted Low-Cost Polarimeter," in IEEE Access, vol. 8, pp. 183965-183971, 2020, doi: 10.1109/ACCESS.2020.3009524.
[12] J. Vojtěch, et al, "CLONETS-DS – Clock Network Services-Design Study Strategy and innovation for clock services over optical-fibre networks in Europe," in Conference on Lasers and Electro-Optics, J. Kang, S. Tomasulo, I. Ilev, D. Müller, N. Litchinitser, S. Polyakov, V. Podolskiy, J. Nunn, C. Dorrer, T. Fortier, Q. Gan, and C. Saraceno, eds., OSA Technical Digest (Optica Publishing Group, 2021), paper JTu3A.30.
[13] Josef Vojtech, Guy Roberts, Tomas Novak, Michal Spacek, Elisabeth Andriantsarazo, Vladimir Smotlacha, Ondrej Havlis, Tomas Horvath, Rudolf Vohnout, Martin Slapak, Jaroslav Roztocil, Susanne Naegele-Jackson, Domenico Vicinanza, Harald Schnatz, Jochen Kronjaeger, Jacques-Olivier Gaudron, and Krzysztof Turza "GÉANT plans towards fibre infrastructure for the distribution of time and frequency throughout Europe", Proc. SPIE 13144, Infrared Remote Sensing and Instrumentation XXXII, 131440N (3 October 2024); https://doi.org/10.1117/12.3027229